\documentstyle[prl,aps,epsfig,twocolumn]{revtex}

\begin{document}
\draft

\twocolumn[\hsize\textwidth\columnwidth\hsize\csname@twocolumnfalse\endcsname
\title{Quasi-spin wave quantum memories with dynamic symmetry}
\author{C. P. Sun $^{1,a,b}$, Y. Li $^{1}$ and X. F. Liu $^{1,2}$}
\address{$^{1}$Institute of Theoretical Physics, the Chinese Academy of Science,
 Beijing, 100080, China\\
 $^{2}$Department of Mathematics, Peking University, Beijing, China}
\maketitle
\begin{abstract}
For the two-mode exciton system formed by the quasi-spin wave
collective excitation of many $\Lambda$ atoms fixed at the lattice
sites of a crystal, we discover a dynamic symmetry depicted by the
semi-direct product algebra $SU(2)\overline{\otimes} h_2 $ in the
large $N$ limit with low excitations. With the help of the
spectral generating algebra method, we obtain a larger class of
exact zero-eigenvalue states adiabatically interpolating between
the initial state of photon-type and the final state of quasi-spin
wave exciton-type. The conditions for the adiabatic passage of
dark states are shown to be valid, even with the presence of the
level degeneracy. These theoretical results can lead to propose
new protocol of implementing quantum memory robust against quantum
decoherence.
\end{abstract}
 ]

Recent progresses in quantum information have stimulated the
development of new concept technologies, such as quantum
computation, quantum cryptography and quantum teleportation
\cite{q-infor}. The practical implementation of these quantum
protocols relies on the construction of both quantum memories
(QMEs) and quantum carriers (QCAs) free of quantum decoherence.
While photons can be generally taken as quantum carriers, quantum
memories should correspond to localized systems capable of storing
and releasing quantum states reversibly. Moreover, to control the
coherent transfer of information, there should be a time dependent
mechanism for turning on and turning off the interaction between
QME and QCA at appropriate instants of time.

A single atom in a cavity QED system \cite{cavity-QED} seems to
satisfy the above mentioned requirements for QME using the Raman
adiabatic passage mechanism \cite{STIRAP}. To achieve the {\it
strong coupling} required for a practical QME a
very elegant method has been proposed recently \cite{Lukin00-ent,Fl00-pol,Fl00-OptCom}%
, where ensembles of $\Lambda $-type atoms are used to store and
transfer the quantum information of photons by the collective
atomic excitations through electromagnetically induced
transparency (EIT) \cite{EIT}. Some experiments \cite{lui,group}
have already demonstrated the central principle of this technique
- the group velocity reduction. The recent success in experiment
also shows the power of such an atomic ensemble QME \cite{polzik}
and motivates additional theoretical works \cite{L-qcommun}. But
there still exists the decoherence problem. An ensemble consists
of many moving atoms and atoms in different spatial positions may
experience different couplings to the controlling external fields.
This results in decoherence in quantum information processing
\cite{sun-you}. To avoid the spatial-motion induced decoherence,
one naturally considers the case that each $\Lambda$ atom is fixed
at a lattice site of a crystal. The most recent experiment of the
ultraslow group velocity of light in a crystal of
$Y_{2}S_{i}O_{5}$ \cite{exp-solid} proposes the possibility of
implementing robust quantum memories by utilizing the solid state
exciton system.

In this letter, we study a system consisting of the quasi-spin
wave collective excitations of many $\Lambda$-type atoms. In this
system, most spatially-fixed atoms stay in the ground state, the
two quasi-spin collective excitations to two excited states behave
as two types of bosons and thus a two mode exciton system forms.
We prove that in the large $N$ limit with low excitations, this
excitonic system possess a hidden dynamic symmetry described by
the semi-direct product $SU(2) \overline{\otimes }h_{2}$ of
quasi-spin $SU(2)$ and the exciton Heisenberg-Weyl algebra
$h_{2}$. With the help of the spectrum generating algebra theory
\cite{algebra} based on $SU(2)\overline{\otimes }h_{2}$, we can
construct the eigen-states of the two mode exciton-photon system
including the collective dark states as a special class. Since the
external classical field is controllable, the quantum information
can be coherently transferred from the cavity photon to the
exciton system and {\it vice versa}. Therefore, the two mode
quasi-spin wave exciton system can serve as a robust quantum
memory.

The model system we consider consists of a crystal with $N$
lattice sites as shown in Fig. 1. There are $N$ 3-level atoms of
$\Lambda $-type with the excited state $|a\rangle $, the relative
ground state $|b\rangle $ and the meta-stable lower state
$|c\rangle $. They interact with two single-mode optical fields.
The transition from $|a\rangle $ to $|b\rangle $ of each atom is
approximately resonantly coupled to a quantized radiation mode
(with coupling constant $g$ and annihilation operator $a$), and
the transition from $|a\rangle $ to $|c\rangle $ is driven by an
exactly resonant classical field of Rabi-frequency $\Omega$. In
recent years, for the similar exciton system in a crystal slab
with spatially fixed two level atoms, both quantum decoherence and
fluorescence process have been extensively studied \cite{sun-liu}.

For convenience we introduce the notation ${\bf j=(}
a_{x}j_{x},a_{y}j_{y},a_{z}j_{z})$ to denote the position of the ${\bf j-}%
th$ site where $a_{u}$ is the length of the crystal cell along $u-$%
direction and $j_{u}=0,1,2,\cdots$, $N_{u}$ for $u=x,y,z$. Then
the quantum dynamics of the total system is described by the
following Hamiltonian in the interaction picture:
\begin{eqnarray}
H &=&ga\sum_{{\bf j}=1}^{N}\exp (i{\bf K}_{ba}\cdot {\bf j})\sigma _{ab}^{%
{\bf j}}  \nonumber \\
&&+\Omega \sum_{j=1}^{N}\exp (i{\bf K}_{ca}\cdot {\bf j})\sigma _{ac}^{{\bf j%
}}+h.c.,
\end{eqnarray}%
where $N=N_{x}N_{y}N_{z}$, and ${\bf K}_{ba}$ and ${\bf K}_{ca}$
are respectively the wave
vectors of the quantum and classical light fields. The quasi-spin operators $%
\sigma _{\alpha \beta }^{{\bf j}}=|\alpha \rangle _{{\bf
jj}}\langle \beta |$ ($\alpha ,\beta =a,b,c$) for $\alpha \neq
\beta $ describe the transition between the levels of $|a\rangle
$, $|b\rangle $ and $|c\rangle$.

\begin{figure}[h]
\begin{center}
\includegraphics[width=4.5cm,height=5cm]{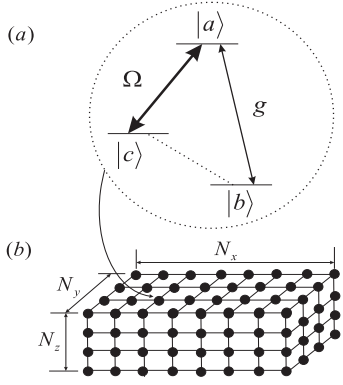}
\vspace{0.3cm}
 \caption{ Configuration of the proposed quantum memory with
$\Lambda $-type atoms. (a) located at lattice sites of crystal.
(b) resonantly coupled to a control classical field and a
quantized probe field. }
\end{center}
\end{figure}

The function of quantum memory is understood in terms of its
quantum state mapping technique. This motivates us to define the
basic quantum states according to the concrete form of the
interaction. We denote by $|v\rangle =|b_{1},b_{2},\cdots
,b_{N}\rangle$ the collective ground
state with all $N$ atoms staying in the same single particle ground state $%
|b\rangle$. It is obvious that, from the ground state $|v\rangle
$, the first order and second order perturbations of the
interaction can create the one exciton quasi-spin wave states
$|1_{a}\rangle$ and $|1_{c}\rangle$:
\begin{equation}
|1_{s}\rangle =\frac{1}{\sqrt{N}}\sum_{{\bf j}=1}^{N}e^{i{\bf
K}_{bs}\cdot {\bf j}}|b,b,\cdots ,\stackrel{{\bf
j-}th}{\overbrace{s}},\cdots ,b\rangle,
\end{equation}
for $s=a,c$ respectively. The wave vector ${\bf K}_{bc}={\bf K}_{ba}-{\bf K%
}_{ca}$ is introduced to depict the second order transition process from $%
|b\rangle $ to $|c\rangle $ as shown in Fig. 2(a). Its collective
effect can be described by operator
\begin{equation}
C=\frac 1{\sqrt{N}}\sum_{{\bf j}=1}^Ne^{-i{\bf K}_{bc}\cdot {\bf
j}}\sigma _{bc}^{{\bf j}},
\end{equation}
which gives $|1_c\rangle \equiv C^{\dagger }|v\rangle $.
Correspondingly, the collective excitation from $|b\rangle $ to
\begin{figure}[h]
\begin{center}
\includegraphics[width=7cm,height=3.5cm]{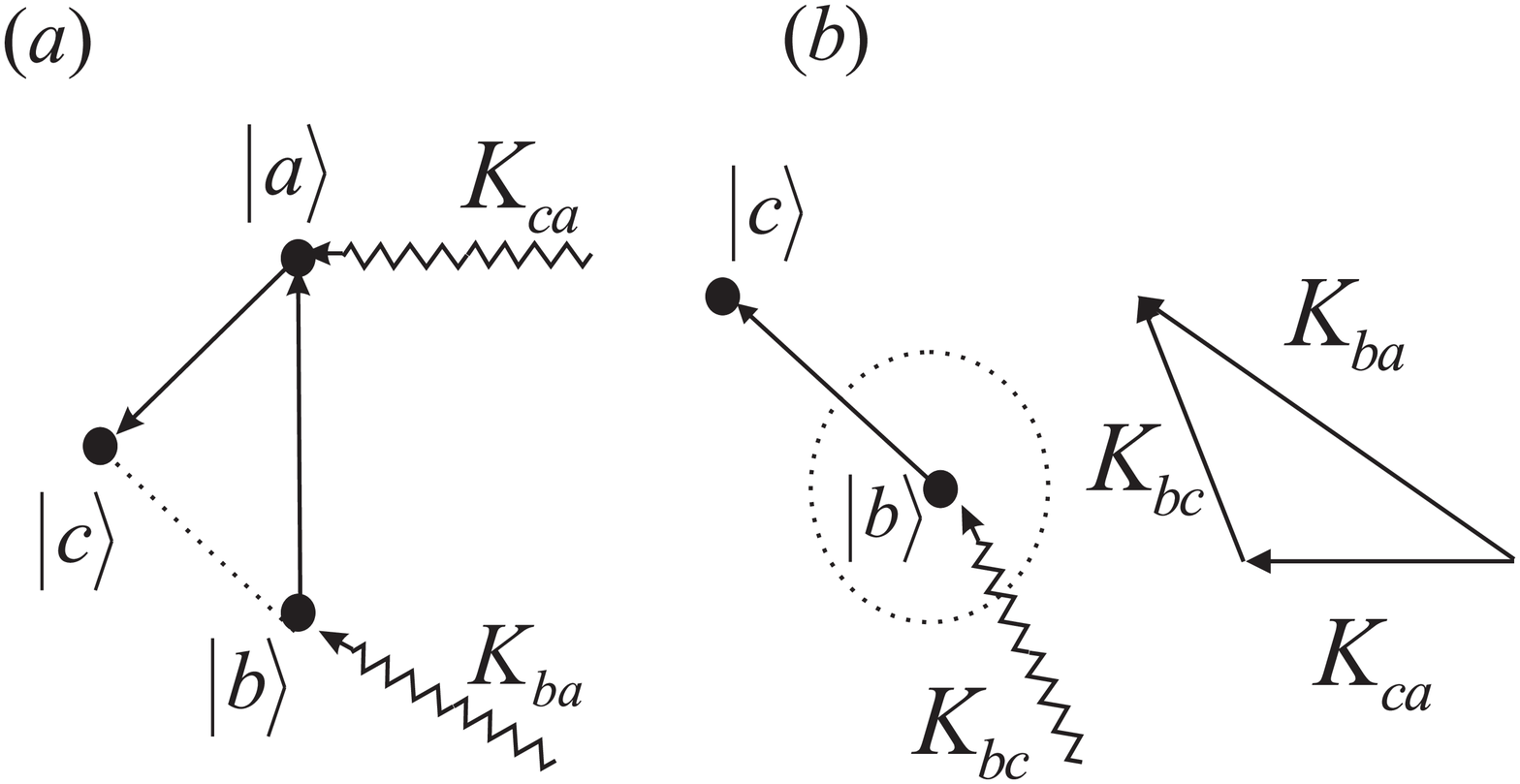}
\end{center}
\vspace{0.3cm}

\caption{Illustration of the second order process $|b\rangle \rightarrow $ $%
|a\rangle \rightarrow $ $|c\rangle $ induced by the classical and
quantized lights.}
\end{figure}

$|a\rangle $ is described by
\begin{equation}
A=\frac{1}{\sqrt{N}}\sum_{{\bf j}=1}^{N}e^{-i{\bf K}_{ba}\cdot {\bf j}%
}\sigma _{ba}^{{\bf j}},
\end{equation}%
which gives $|1_{a}\rangle \equiv A^{\dagger }|v\rangle$. In the
large $N$ limit with the low excitation condition that there are
only a few atoms occupying $|a\rangle $ or $|c\rangle $
\cite{q-defor}, the two mode quasi-spin wave excitations behave as
two bosons since in this case they satisfy the bosonic commutation
relations $[A,A^{\dagger }]=1$, $[C,C^{\dagger }]=1$. In the same
limit, it is worth to point out that $[A,C]=0$ and $[A,C^{\dagger
}]=-T_{-}/N\longrightarrow 0$, thus these quasi-spin wave low
excitations are independent of each other.

In terms of these two mode exciton operators, the interaction Hamiltonian
reads
\begin{equation}
H=g\sqrt{N}aA^{\dagger }+\Omega T_{+}+h.c.,
\end{equation}%
where the collective operators
\begin{equation}
T_{-}=\sum_{{\bf j}=1}^{N}e^{-i{\bf K}_{ca}\cdot {\bf j}}\sigma _{ca}^{{\bf j%
}}, T_{+}=(T_{-})^{\dagger },
\end{equation}%
generate the $SU(2)$ algebra together with the third generator $T_{3}=\sum_{{\bf j}=1}^{N}(\sigma _{aa}^{%
{\bf j}}-\sigma _{cc}^{{\bf j}})/2$. It is very interesting to
observe that the exciton operators and the $SU(2)$ generators span
a larger Lie algebra. By a straightforward calculation we obtain
\begin{equation}
\lbrack T_{-},C]=-A,[T_{+},A]=-C.
\end{equation}%
Denote by $h_{2}$ the Lie algebra generated by $A$, $A^{\dagger}$, $C$, and $%
C^{\dagger}$. It then follows that $[SU(2),h_{2}]\subset h_{2}$.
This means
that in the large $N$ limit with the low excitation condition the operators $%
A$, $A^{\dagger}$, $C$, $C^{\dagger}$, $T_{3}$, $T_{\pm}$ and the
identity $1$ span a semidirect product Lie algebra
$SU(2)\overline{\otimes }h_{2}$. In the following discussion we
will focus on this case except otherwise explicitly specified.
Since the Hamiltonian $H$ can be expressed as a function of the
generators of $SU(2)\overline{\otimes }h_{2}$, one says that the
two-mode exciton system possesses a dynamic symmetry governed by
the dynamic \textquotedblleft group\textquotedblright\ (or dynamic algebra) $%
SU(2)\overline{\otimes }h_{2}$. The discovery of this dynamic
symmetry leads us, by the spectrum generating algebra method
\cite{algebra}, to find $H-$ invariant subspaces, in which one can
diagonalize the Hamiltonian easily.

To that cause, we define
\begin{equation}
D=a\cos \theta -C\sin \theta
\end{equation}%
with $\theta(t)$ satisfying $\tan \theta
(t)=\frac{g\sqrt{N}}{\Omega (t)}$. It mixes the electromagnetic
field and collective atomic excitations of quasi spin wave.
Evidently, $[D,D^{\dagger }]=1$ and $[D,H]=0$. Thus the
Heisenberg-Weyl group $h$ generated by $D$ and $D^{\dagger }$ is a
symmetry
group of the two-mode exciton-photon system. We introduce the state $|{\bf 0}%
\rangle =|v\rangle \otimes |0_{l}\rangle $ where \ $|0_{l}\rangle
$ is the vacuum of the electromagnetic field, we find $D|{\bf
0}\rangle =0$ and it is an eigen-state of $H$ with zero
eigen-value. Consequently, a degenerate class of eigen-states of
$H$ with zero eigen-value can be constructed naturally as follows:
\begin{equation}
|d_{n}\rangle =[n!]^{-1/2}D^{\dagger n}|{\bf 0}\rangle.
\end{equation}%
Physically, the above dressed state is cancelled by the interaction
Hamiltonian and thus is called a dark state or a dark-state polariton (DSP).
A DSP traps the electromagnetic radiation from the excited state due to
quantum interference cancelling. For the case with an ensemble of free
moving atoms, the similar DSP was obtained in refs. \cite%
{Lukin00-ent,Fl00-pol,Fl00-OptCom} to clarify the physics of the
state-preserving slow light propagation in EIT associated with the existence
of quasi-particles.

Now starting from these dark states $|d_{n}\rangle $, we can use the
spectrum generating algebra method to build other eigenstates for the total
system. To this end we introduce the bright-state polariton operator
\begin{equation}
B=a\sin \theta +C\cos \theta .
\end{equation}%
It is easy to check that $[B,B^{\dagger }]=1$, and $[D,B^{\dagger
}]=[D,B]=0$. Evidently $[A,B]=[A,B^{\dagger }]=0$, this amounts to
the fact that $A$ commutes with $C$ and $C^{\dagger }$ in the
large $N$ limit with low excitations. What is crucial for our
purpose is the commutation
relations $[H,Q_{\pm }^{\dagger }]=\pm \epsilon Q_{\pm }^{\dagger }$ for $%
Q_{\pm }=\frac{1}{\sqrt{2}}(A\pm B)$ where $\epsilon =\sqrt{g^{2}N+\Omega
^{2}}$. Thanks to these commutation relations we can construct the
eigen-states
\begin{equation}
|e(m,k;n)\rangle =[m!k!]^{-1/2}Q_{+}^{\dagger m}Q_{-}^{\dagger
k}|d_{n}\rangle,
\end{equation}%
as the dressed states of the two-mode exciton system. The corresponding
eigen-values are
\begin{equation}
E(m,k)=(m-k)\epsilon ,m,k=0,1,2\cdots.
\end{equation}%
We notice that for each given pair of indices $(m,k)$,
$\{|e(m,k;n)\rangle |n=0,1,2,\cdots \}$ defines a degenerate set
of eigen-states. Physically the spectral structure of the dressed
two-mode exciton system resembles that of a two mode harmonic
oscillator, but its energy level number is finite and each energy
level possesses a very large degeneracy.

The above equations show that there exists a larger class ${\bf S}%
:\{|e(m,m;n)\rangle \equiv |d(m,n)\rangle |$ $\ m,n=0,1,2,\cdots
\}$ of states of zero-eigen-value $E(m,m)=0$. They are constructed by acting $%
Q_{+}^{\dagger }Q_{-}^{\dagger }$ \ $m$ times on $|d_{n}\rangle $
:
\begin{equation}
|d(m,n)\rangle =\sum_{k=0}^{m}\frac{A^{\dagger 2(m-k)}B^{\dagger 2k}}{%
2^{m}(m-k)!k!}|d_{n}\rangle.
\end{equation}%
This larger degeneracy is physically rooted in the larger symmetry group $%
h_{2}$ generated by $Q_{+}^{\dagger }Q_{-}^{\dagger }$ and $Q_{+}Q_{-}$
together with $D$ and $D^{\dagger }$. The original quantum memory defined by $%
\left\{ |d_{n}\rangle =|d(0,n)\rangle \right\} $ in ref. \cite
{Lukin00-ent,Fl00-pol,Fl00-OptCom} actually is a special subset of
the larger class.

Now we consider whether these states of zero-eigen-value can work
well as a quantum memory by the adiabatic manipulation \cite
{Lukin00-ent,Fl00-pol,Fl00-OptCom}. The quantum adiabatic theorem \cite%
{prd,zee} for degenerate cases tells us that, under the adiabatic
condition
\begin{equation}
|\frac{\langle e(m,k;n)|\partial _{t}|d(m,n)\rangle }{E(m,k)}|\sim \frac{g%
\sqrt{N}\stackrel{.}{|\Omega (t)|}}{\epsilon ^{3}}\ll 1,
\end{equation}
the adiabatic evolution of any degenerate system will keep itself
within the block ${\bf S}$ of dark states with the same
instantaneous eigen-value $0$. However, it does not forbid
transitions within states in this block ${\bf S}$, such as those
between $\left\{ |d_{n}\rangle =|d(0,n)\rangle \right\}$ and
$\{|d(m,n)\rangle (m\neq 0)\}$. So it is important to consider
whether there exists any dynamic mechanism to forbid such
transitions. Actually this issue has been uniformly ignored in all
previous studies even for the degenerate set $\left\{
|d_{n}\rangle \right\}$.

We can generally consider this problem by defining the zero-eigenvalue subspaces $%
{\bf S}^{[m]}:\{|d(m,n)\rangle |$ $\ n=0,1,2,\cdots \}$, ${\bf S}^{[0]}={\bf S}%
$. The complementary part of the direct sum ${\bf DS}={\bf S}^{[0]}\oplus $ $%
{\bf S}^{[1]}\oplus\cdots$ of all dark state subspaces is ${\bf
CS=}\{|e(m,k,n)\rangle |k\neq m,n=0,1,2,\cdots \}$ in which each
$|e(m,k,n)\rangle $ has non-zero eigenvalue. Any state $|\phi
^{\lbrack m]}(t)\rangle =\sum_{n}c_{n}^{[m]}(t)|d(m,n)\rangle $ in ${\bf S}%
^{[m]}$ evolves according to
\begin{equation}
i\frac{d}{dt}c_{n}^{[m]}(t)=\sum_{m^{\prime },n^{\prime
}}D_{mn}^{m^{\prime }n^{\prime }}c_{n^{\prime }}^{[m^{\prime
}]}(t)+F[{\bf CS}],
\end{equation}%
where $F[{\bf CS}]$, which can be ignored under the adiabatic
conditions \cite{prd,zee}, represents a certain functional of the
complementary states and $D_{mn}^{m^{\prime }n^{\prime}}=-i\langle
d(m^{\prime },n^{\prime
})|\partial _{t}|d(m,n)\rangle $. Considering $\partial _{\theta }B=D$ and $%
\partial _{\theta }D=-B,$ we have $D_{mn}^{m^{\prime }n^{\prime
}}=-i\stackrel{.}{\theta }\langle d(m^{\prime },n^{\prime
})|\partial _{\theta }|d(m,n)\rangle $. The equation about
$\partial _{\theta }|d(m,n)\rangle $ contains 4 terms: $|e(m,m\mp
1;n\pm 1)\rangle $ and $|e(m\mp 1,m;n\pm 1)\rangle $. This implies
the exact result\ $\langle d(m^{\prime },n^{\prime })|\partial
_{\theta }|d(m,n)\rangle $ $=0$, showing there is indeed no mixing
among the dark states during the adiabatic evolution. Viewed from
physical aspect, this can also be understood as the adiabatic
change of external
parameters do not lead the system to enter into the complementary space $%
{\bf CS}$. Notice that only for the non-adiabatic evolution, will
the non-zero matrix elements $\langle e(m^{\prime },k,n^{\prime
})|\partial _{\theta }|d(m,n)\rangle$ will be  a cause for state
mixing. The same physics has been considered in the context of the
Abelianization of the non-Abelian gauge structure induced by an
adiabatic process \cite{prd,zee}. This argument gives a necessary
theoretical support for the practical realization of the original
scheme of quantum memory by Fleischhauer, Lukin and their
collaborators \cite{Lukin00-ent,Fl00-pol,Fl00-OptCom}.

Based on the above consideration, we thus claim that, for each
fixed $m\neq 0$, each subspace ${\bf S}^{[m]}$ can work formally
as a quantum memory different from that in ref.
\cite{Lukin00-ent,Fl00-pol,Fl00-OptCom}. We introduce the notation
\begin{equation}
|{\bf A,P,}m\rangle =\frac{1}{2^{m}m!}(A^{\dagger 2}-P^{\dagger 2})^{m}|{\bf %
0}\rangle
\end{equation}%
for ${\bf P=a,C}$. Both the initial state $|d(m,n)\rangle
|_{\theta =0}=|{\bf A,C,}m\rangle \otimes |n\rangle _{L}$ and the
final state $|d(m,n)\rangle |_{\theta =\pi /2}=
|n\rangle _{C}\otimes $ $(-1)^{n}$\ $%
|{\bf A,a,}m\rangle$ have factorization structure. Thus we can use
the general initial state $|s(0)\rangle =\sum_{n}c_{n}\,|n\rangle
_{L}$ of single-mode light to record quantum information and
prepare the exciton in a paired state $|{\bf A,C,}m\rangle$. When
one rotates the mixing angle $\theta$ from 0 to $\pi /2$ by
changing the coupling strength $\Omega (t)$ adiabatically, the
total system will reach the final state $|S(t)\rangle
=(\sum_{n}c_{n}\,|n\rangle _{C})\otimes |{\bf A,a,}m\rangle $ with the $c$%
-mode quasi-spin wave decoupling with the other parts. From the
viewpoint of quantum measurement the decoding process is then to
average over the states of photon and $A$-exciton and to obtain
the pure state density matrix $\rho
_{C}=\sum_{n,m}c_{n}c_{m}^{\ast }\,|n\rangle _{cc}\langle m|$,
which is the same as that for the initial photons. Therefore, the
above discussion suggests a new protocol of storing quantum
information when the decay  of excited state is enough small
during adiabatic manipulation.

Before concluding, we would like to address that, the individual
atoms in the generalized states $|d(m,n)\rangle $ have excited
state components and therefore $|d(m,n)\rangle $ is not totally
dark in practice. If the excited state decays faster, the
generalized states $|d(m,n)\rangle$ would also decay during slow
adiabatic manipulation. This metastable nature would leads to an
undesirable effect for memory application. We also point out that
the present treatment is only valid for the low density excitation
regime where the bosonic modes of the quasi-spin wave excitations
can be used effectively. Therefore the above  down Fock state
formally written as $A^{\dagger m}C^{\dagger n}|{\bf 0}\rangle $
doesn't make sense when $m$ or $n$ is large. By the mathematical
duality, the situation with extremely-high excitation can be dealt
with in a similar manner. In fact, the serious difficulty only
lies in the region  where the excitation is neither very low nor
very high. In that case, it turns out that the boson commutation
relation of the excitaton operators must be modified, for example,
to the $q$-deformed one ($q=1-O(\frac{1}{N}))$ \cite{q-defor}.
Physically this modification will cause quantum decoherence of the
collective degrees of freedom in the exciton system. Finally we
emphasize that, though in our model system assumed to be located
at regular lattice sites as in a crystal, our results (at least in
mathematical formulation ) remain valid for an ensemble of atoms
with random spatial positions, provided that we can ignore the
kinetic energy terms (of the center of mass motion) of the atom.
It seems that the ensemble of free atoms can function as quantum
memory of same kind. However the strict treatment of the atomic
ensemble based quantum memory should include the kinetic energy
terms of the atom center of mass. The momentum transfer of atomic
center of mass can induce additional quantum decoherence
\cite{sun-you}. In our present protocol, this decoherence effect
is partly overcome by fixing atoms at lattice sites and thus
neglecting the kinetic energy terms.

We acknowledge the support of the CNSF (grant No.90203018) and the
knowledged Innovation Program (KIP) of the Chinese Academy of
Sciences and the National Fundamental Research Program of China
with No.001GB309310. We also sincerely thank L. You and H. Pu for
helpful discussions.

\end{document}